\documentclass[12pt,epsfig]{article}

\usepackage{amssymb,epsfig}

\newcommand{\be}{\begin{equation}}

\newcommand{\ee}{\end{equation}}

\newcommand{\bea}{\begin{eqnarray}}

\newcommand{\eea}{\end{eqnarray}}

\newcommand{\kt}{\vec{k}}
\newcommand{\p}{\vec{p}}

\begin{document}

\begin{titlepage}

\begin{flushright}
\begin{tabular}{l}
 CPHT RR 059.0602\\
 hep-ph/0209300
\end{tabular}
\end{flushright}
\vspace{1.5cm}

\begin{center}

{\LARGE \bf
Probing chiral-odd GPD's in diffractive electroproduction of two vector
mesons}

\vspace{1cm}

{\sc D.Yu.~Ivanov}${}^{1,2}$,
{\sc B.~Pire}${}^{3}$,
{\sc L.~Szymanowski}${}^{3,4}$ and
{\sc O.V.~Teryaev}${}^{3,5}$
\\[0.5cm]
\vspace*{0.1cm} ${}^1${\it
   Institut f{\"u}r Theoretische Physik, Universit{\"a}t
   Regensburg, \\ D-93040 Regensburg, Germany
                       } \\[0.2cm]
\vspace*{0.1cm} ${}^2$ {\it
Institute of Mathematics, 630090 Novosibirsk, Russia
                       } \\[0.2cm]
\vspace*{0.1cm} ${}^3$ {\it
CPhT, {\'E}cole Polytechnique, F-91128 Palaiseau, France\footnote{
  Unit{\'e} mixte C7644 du CNRS.}
                       } \\[0.2cm]
\vspace*{0.1cm} ${}^4$ {\it
 So{\l}tan Institute for Nuclear Studies,
Ho\.za 69,\\ 00-681 Warsaw, Poland
                       } \\[0.2cm]
\vspace*{0.1cm} ${}^5$ {\it
Bogoliubov Lab. of Theoretical Physics, JINR, 141980 Dubna, Russia
                       } \\[1.0cm]
{\it
 }
\vskip2cm
\end{center}
We consider the electroproduction of two vector mesons
with a large rapidity gap between them  on a nucleon target
 in the process
$ \gamma ^* N \to \rho_1 \rho_2 N'$. We calculate the Born term
for this
process within the
collinear factorization framework.
The resulting scattering amplitude may be represented
as a convolution
of an impact factor describing the $\gamma ^* \to \rho_1$ transition
and an amplitude describing the $N\to \rho_2 N'$ transition.
The latter amplitude is analogous to deeply virtual
electroproduction of a meson, the
virtual photon being replaced by two gluon (Pomeron) exchange .
The long distance
part of this amplitude is described by
 Generalized Parton Distributions (GPD) and meson light-cone distributions.
The
selection of a transversely polarized vector meson $\rho_2$
provides the first feasible selective access to chiral-odd GPD.
\vskip1cm

\vspace*{1cm}

\end{titlepage}

{\large \bf 1.~~}
Generalized Parton Distributions (GPDs) are the non-perturbative
objects encoding the information about the quark and gluon proton
structure in the most complete way \cite{GPD}. While the chiral even GPD
may be probed in various hard exclusive processes, no single one has yet
be proven to be sensitive to  chiral-odd GPDs \cite{COLLINS},
\cite{COGPD}.

In the massless quark limit, the chiral-odd functions may appear
only in pairs in a non-vanishing scattering amplitude, so that chirality
flip encoded in one of them is compensated by another. The natural probe
for the forward chiral-odd distributions is the Drell-Yan process,
containing the convolution of chiral-odd distributions of quark and
antiquark \cite{tra}.
Its nonforward analog is provided by the hard exclusive
production of a transversely polarized vector meson, where the
quark transversity distribution in the nucleon  is
substituted by the chiral-odd GPD and the antiquark one by the
meson distribution amplitudes (DA)
However, the simplest realization of this idea \cite{COLLINS}, namely the
hard
exclusive electroproduction of a transversely polarized vector meson,
results in a zero contribution \cite{DGP}. Formally, this comes
>from the vanishing value of the expression
$\gamma^\mu \sigma_{\alpha \beta}\gamma_\mu $,
appearing after the summation over the polarizations
of the virtual gluon, which is required to transfer the hard
momentum transfer in the case of collinear GPD and DA. This formal
argument is supported by the consideration of angular momentum
conservation in the collinear kinematics.

In the present paper we suggest another process
which  allows to avoid these effects by "substituting" to the
virtual photon a hard two gluon exchange, {\it i.e.}  a perturbative
Pomeron (${\cal P}$) in
the lowest
order,
coming from a photon/meson
transition.
 Let us consider a generic process
\be
A N \to B M N'
\label{process}
\ee
shown in Fig. 1 of scattering
of a particle $A$, e.g.  being a virtual or real photon, on a
nucleon $N$, which leads via two gluon exchange to the production
of  particle $B$ (e.g. vector meson or photon) separated by a large
rapidity gap from another produced meson $M$ and the scattered nucleon $N'$.
We consider the kinematical region where the rapidity gap between $M$ and
$N'$ is much smaller than the one between
$B$ and $M$, that is the energy of the system ($M\; -\; N'$) is smaller
than the energy of the system ($B\; -\; M$) but still large enough to
justify our approach  (in particular much larger
than baryonic resonance masses).

We show that in  such kinematical
circumstances
the Born term for this
process is calculable consistently within
the collinear factorization method. The final result is represented as an
integral (over  the
longitudinal momentum fractions of the quarks)  of
the product of two amplitudes: the first one
 describing
the transition $A \to B$ via two gluon exchange and
the second one  describing
the subprocess
${\cal P}\;N\;\to \;M\;N'$ which is
closely related to the electroproduction process $\gamma^*\,N \to M\,N'$
where  collinear factorization
theorems allow to separate  the long distance dynamics  expressed
through the
GPDs from a perturbatively calculable coefficient function. The hard scale
appearing in the process in Fig. 1 is supplied by the
relatively large  momentum transfer
$p^2$ in the two gluon channel, i.e. by the virtuality of the Pomeron.

Such a process is a representative of a new class of hard
reactions whose QCD description within the collinear factorization scheme
involves
in the described above kinematics
the impact factor $J$ appearing naturally
in Regge-type perturbative description
based on the BFKL evolution and it involves
the collinear
distributions (e.g. F), whose evolutions are governed by DGLAP-ERBL
equations.

In this place we would like to mention Ref.~\cite{Freund} in which the
factorization theorem was proven for  electroproduction of multiple mesons
of small invariant mass. Consequently, this bulk of mesons is described in
\cite{Freund} by
 one generalized distribution amplitude. This should be contrasted with the
case which we consider
below where the two produced mesons have a large invariant mass,
which leads to a different, yet unproven to all orders, factorization theorem.

In order to make our discussion as clear as possible we shall
consider first a reference
process with all longitudinally polarized vector particles
\be
\label{2mesongen}
\gamma^*_L (q)\;\; N (p_2) \to  \rho_L^0(q_\rho)\;\; \rho_L^+(p_\rho)
N'(p_{2'})\;,
\ee
which involves the emission of two gluons
 in the $\gamma^*_L \to \rho_L$ transition and
which is more founded theoretically from the point of view
of the collinear factorization.
We choose a charged vector meson $\rho^+$ to select quark antiquark exchange
with the nucleon line.
We shall show below
that the collinear factorization holds
at least in the Born
approximation of (\ref{2mesongen}) and we shall derive  the
resulting scattering
amplitude. At the same time it will be clear that all steps of derivation
can be
immediately applied to the
description of a whole family of processes, in particular
those involving the chiral-odd GPD,
 e.g. for
\be
\label{2mesontr}
\gamma^*_L (q)\;\; N (p_2) \to  \rho_L^0(q_\rho)\;\; \rho_T^+(p_\rho)
N'(p_{2'})\;,
\ee
which has been the main motivation for the present studies.
We finish our paper by discussing some other processes which could be useful
for studies of the
transversity
and present final conclusions.

%
\begin{figure}[t]
\centerline{\epsfxsize8.0cm\epsffile{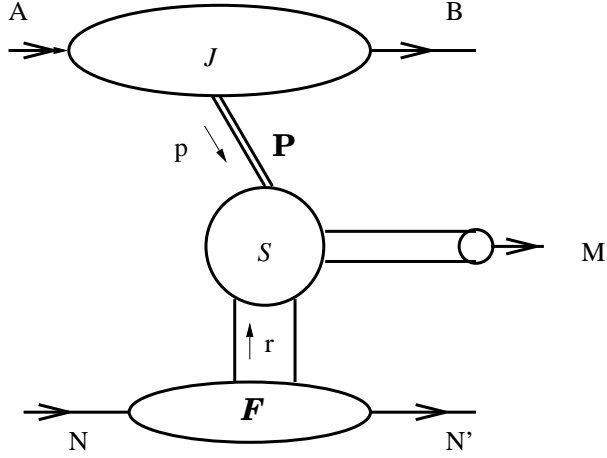}}
\caption[]{\small
Factorization of the process $A\;N \to \;B\;M\;N'$ in the asymmetric
kinematics
discussed in the text. ${\bf P}$ is the hard Pomeron.
 }
\label{fig:1}
\end{figure}
%

\vskip.1in
{\large \bf 2.~~}
Let us first summarize the details of the kinematics of the process
(\ref{2mesongen}).
We introduce two light-like Sudakov vectors
$p_{1/2}$. The momenta are parametrized as follows~:
\bea
&& q^\mu = p_1^\mu -
\frac{Q^2}{s}p_2^\mu\;,\;\;\;\;q^2=-Q^2\,,\;\;\;\;s=2(p_1p_2)\nonumber \\
&& q_\rho^\mu = \alpha p_1^\mu +
\frac{\p^{\;2}}{\alpha s}p_2^\mu + p_\perp^\mu \;,\;\;\;\;\;p_\perp^2 =
-\p^{\;2}
\nonumber \\
&& p_\rho^\mu = \bar \alpha p_1^\mu + \frac{\p^{\;2}}{\bar \alpha s}p_2^\mu
- p_\perp^\mu \;,\;\;\;\;\;\bar \alpha \equiv 1-\alpha \nonumber \\
&& p_{2'}^\mu = p_2^\mu (1- \zeta)
\label{Sud}
\eea
where $\zeta$ is the skewedness parameter  which can be written in
terms of
the two meson invariant mass
\be
s_1 = (q_\rho + p_\rho)^2 = \frac{\p^{\;2}}{\alpha \bar \alpha}
\label{s1}
\ee
and the photon virtuality $Q^2$ as
\be
\zeta = \frac{1}{s}\,\left(Q^2 + s_1  \right)\,.
\label{zeta}
\ee
The $\rho^+(p_\rho)-$meson - target invariant mass equals
\be
s_2 = (p_\rho + p_{2'})^2 = s\, \bar \alpha\,\left(1-\zeta  \right)\,.
\label{s2}
\ee
The kinematical limit with a large rapidity gap between the two mesons in the
final state is
obtained by demanding that $s_1$ is very large, being of the order of $s$
\be
s_1 =s\, \zeta\,,\;\;\;\;s_1 \gg Q^2,\,\,\p^{\;2}\,,
\label{s1gap}
\ee
whereas $s_2$ is kept constant but  large enough to justify the use
of  perturbation theory in the
collinear subprocess ${\cal P} N \to \rho^+_L N'$ and the application of
the GPD framework
\be
\label{s2gap}
s_2 \to \frac{\p^{\;2}}{\zeta}\,\left(1-\zeta  \right) = constant\,.
\ee
In terms of the longitudinal fraction $\alpha$ the limit
with a large rapidity gap corresponds
to taking the limits
\be
\label{alphagap}
\alpha \to 1\,,\;\;\;\;\;\bar \alpha s_1 \to \p^{\;2}\,,\;\;\;\;\;\;\zeta
\sim
1\,.
\ee
We have choosen the kinematics so that the nucleon gets no transverse
momentum in
the process. Let us however note that in principle one may allow a finite
momentum transfer, small with respect to $|\p|$. This case will involve
additional GPDs in the expressions to follow.

\vskip.1in
Let us repeat in this place that the role of the main hard scale in the
processes under discussion below is played by the virtuality $p^2=-\p^2$,
or by the large momentum transfer
in  the two-gluon exchange channel. If additionally the incoming photon
has non zero, sufficiently large  virtuality $Q^2$,  then the
theoretical description of the processes simplifies even more,
as we can neglect within our approximation
contribution of the hadronic
component of the photon. The case with $Q^2=0$, i.e. the photoproduction
at large momentrum transfer is more complicated and may require to take into
account both the perturbative and the hadronic (non-perturbative)
contributions (see e.g. \cite{larget}).

\begin{figure}[t]
\begin{minipage}[t]{72mm}
\centerline{\includegraphics[scale=0.95]{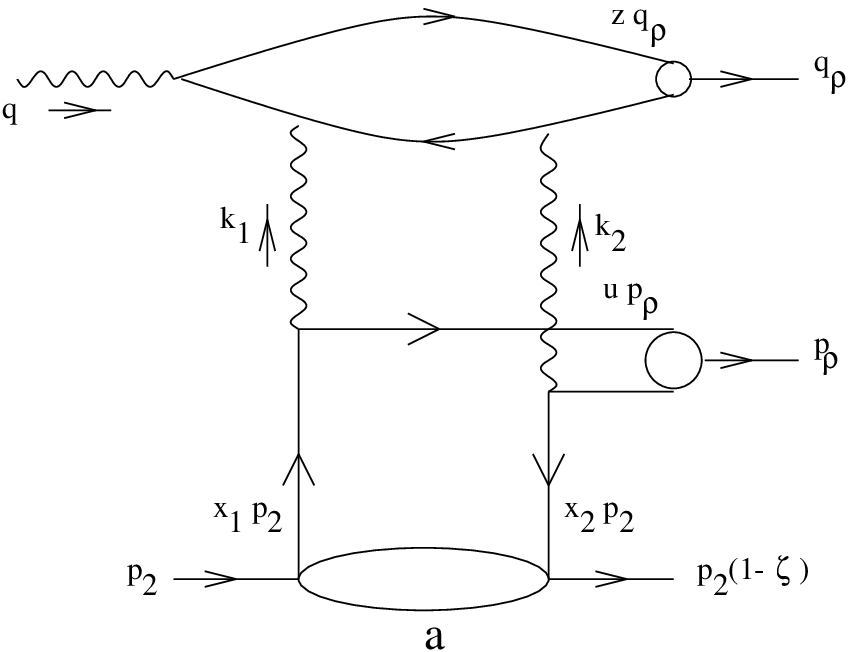}}
\end{minipage}
\hspace{\fill}
\begin{minipage}[t]{72mm}
\centerline{\includegraphics[scale=0.95]{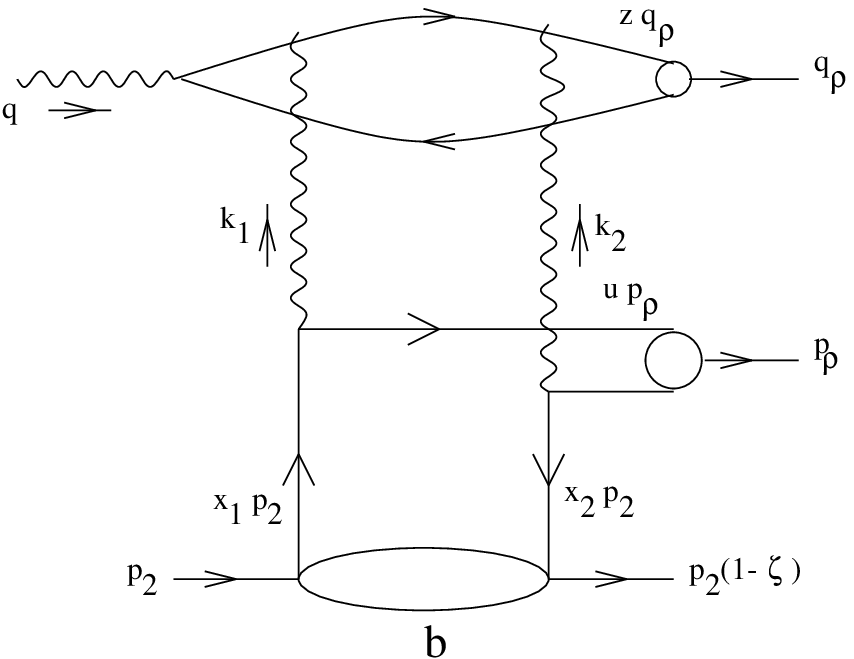}}
\end{minipage}
\\
\vskip.2in
\begin{minipage}[t]{72mm}
\centerline{\includegraphics[scale=0.95]{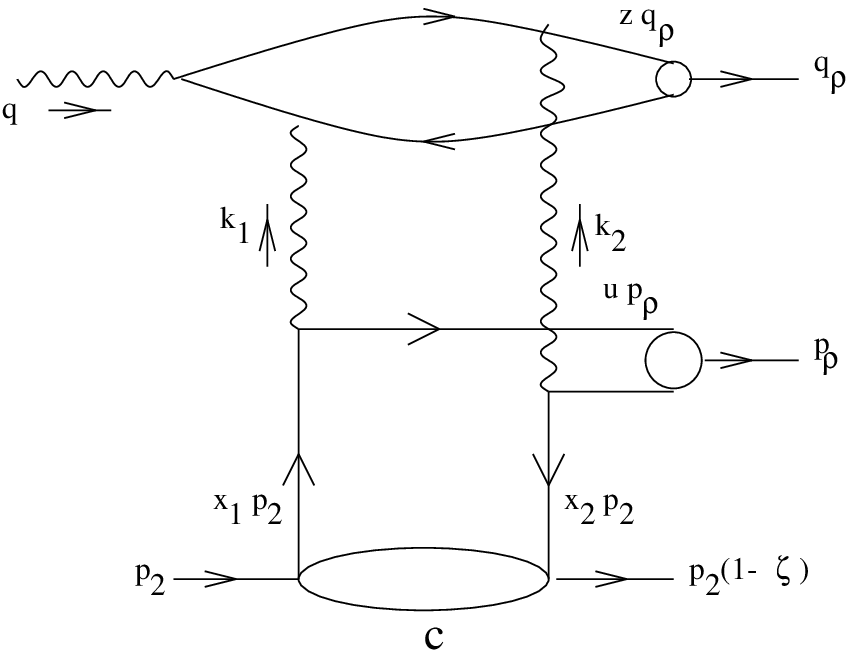}}
\end{minipage}
\hspace{\fill}
\begin{minipage}[t]{72mm}
\centerline{\includegraphics[scale=0.95]{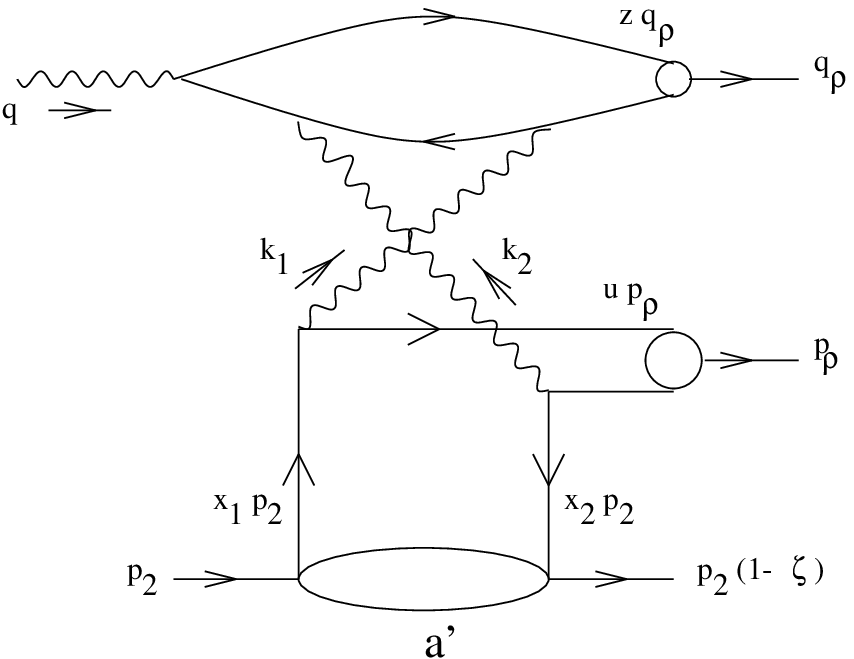}}
\end{minipage}
\\
\vskip.2in
\begin{minipage}[t]{72mm}
\centerline{\includegraphics[scale=0.95]{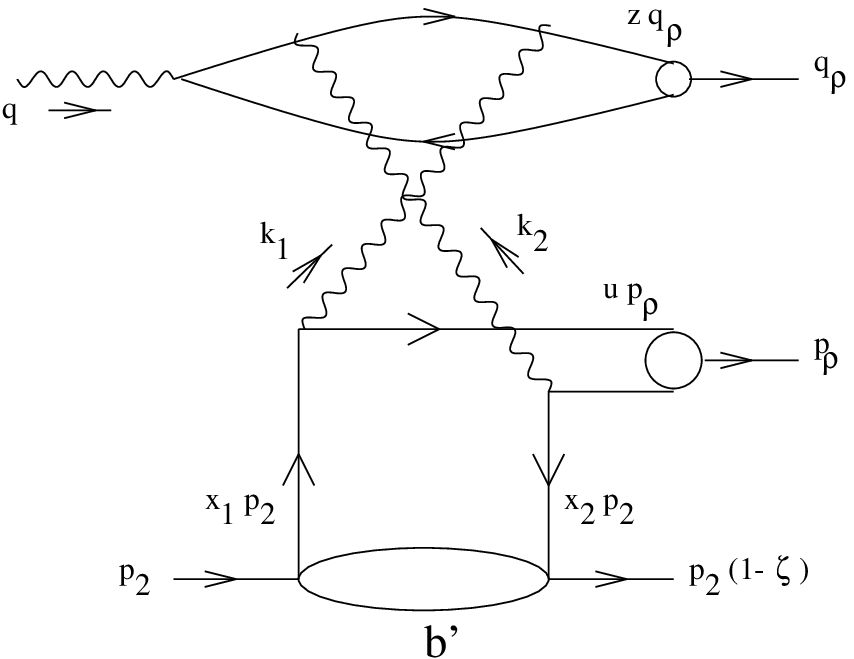}}
\end{minipage}
\hspace{\fill}
\begin{minipage}[t]{72mm}
\centerline{\includegraphics[scale=0.95]{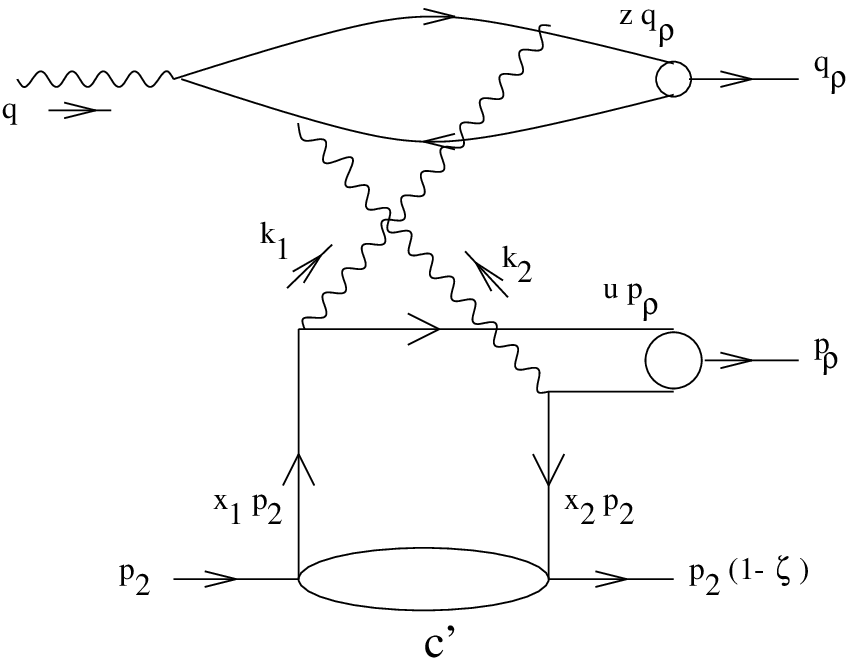}}
\end{minipage}
\caption{{\protect\small Diagrams contributing to the scattering amplitude
in the
Born approximation }}
\label{6d}
\end{figure}

\vskip.1in
{\large \bf 3.~~} We calculate the scattering amplitude ${\cal M}$ of the
process
(\ref{2mesongen})  using the standard collinear factorization
method, i.e. we write it in a form suggested by Fig.~(\ref{fig:1}), as:
\be
\label{fact}
{\cal M} = \sum\limits_{p=q,\bar q}\int\limits_0^1 dz\,\int\limits_0^1
du\,\int\limits_0^1 dx_1
T^p_H(x_1,u,z)\,F^p_\zeta(x_1) \phi_{\rho^+}(u)
\phi_{\rho^0}(z)\;.
\ee
Here $F^p_\zeta(x_1)$ is the generalized (skewed) parton $p$ distribution
in the
target at zero momentum transfer; $x_1$ and $x_2 = x_1 -\zeta$ are (see
Fig.~(\ref{6d})) the momentum
fractions of the emitted and absorbed partons (quarks) of the target,
respectively (as usual the case $x_2 < 0$ is interpreted as an emitted
antiquark).
$\phi_{\rho^+}(u)$ and $\phi_{\rho^0}(z)$ are the distribution
amplitudes of the $\rho^+-$meson and
$\rho^0-$meson, respectively.
$T^p_H(x_1,u,z)$ is
the hard scattering amplitude (the coefficient function).
For clarity of notation we omit in Eq.~(\ref{fact}) the factorization scale
dependence of $T^p_H$, $F^p_\zeta$, $\phi_{\rho^0}$ and $\phi_{\rho^+}$.

Eq.~(\ref{fact}) describes the amplitude in the leading twist
approximation. In other words
all terms suppressed by powers of a hard scale parameter $1/|\p|$ are
omitted.  Within this approximation one neglects (in the physical gauge)
the contributions of the higher Fock states in the meson wave functions
and the many parton correlations (higher twist GPD's) in the proton.
Moreover, one can neglect in the hard scattering amplitude the relative
(with respect to a meson momentum) transverse momenta of constituent
quarks (the collinear approximation). This results in the appearence in
the factorization formula (\ref{fact}) of the distribution amplitudes,
i.e. the usual wave functions depending on the relative transverse
momenta of constituents integrated over these momenta up to the collinear
factorization scale.

The additional simplification appears in the kinematics given by
Eqs.~(\ref{s1gap}-\ref{alphagap}). In this limit one
can consider only the diagrams which involve two gluon exchange between two
mesons, see
Fig.~2. The other contributions (the fermion exchange diagrams) to the
coefficient function $T_H^p$
are known \cite{BFKL} to be suppressed by the power of energy, $\sim
\p^{\;2}/s$, therefore we
will
not discuss them in what follows. We will show that at the same accuracy,
i.e. neglecting
terms $\sim \p^{\;2}/s$, the contribution of gluon exchange diagrams shown
in Fig. 2 is purely
imaginary and involves GPD's in the ERBL region only, $x_1\, <\, \zeta$.
The first property is
of a
general nature, it is valid for any diffractive processes calculated in the
Born approximation
(two gluon
exchange), see \cite{BFKL}. The second one is specific for our process.

In the Born approximation the scattering amplitude $T^q_H(x_1,u,z)$ for the
quark $q$ target
is described by six diagrams, see Fig.~\ref{6d}. They are calculated for the
on-mass-shell quarks carrying the collinear momenta $x_{1,\,2}p_2$.
 The on-mass-shell quark and antiquark entering the $\rho-$mesons
distribution amplitudes $ \phi_{\rho^+}(u)$ and $\phi_{\rho^0}(z)$
carry fractions $u$ and $z$ of the momentum of a corresponding outgoing
mason, $q_\rho$ and
 $p_\rho$, respectively.

We shall show below that for the process (\ref{2mesongen}) the integrals
over $x_1, u, z$ are
convergent which justifies the validity of the factorization formula
(\ref{fact}).

{\large \bf 4.~~} We define
the light-cone distribution amplitudes $\phi_\parallel$ of
$\rho^{0,+}_L-$mesons appearing in
(\ref{2mesongen}) by the matrix elements
\be \label{rho0} \langle 0 |\bar q(0) \gamma^\mu
q(y)| \rho^0_L(q_\rho) \rangle = q_\rho^\mu\, f_\rho \int\limits_0^1
dz\,e^{-iz(q_\rho\,y)}
\phi_\parallel(z) \ee \be \label{rho+L} \langle 0 |\bar q(0) \gamma^\mu
q(y)| \rho^+_L(p_\rho)
\rangle = p_\rho^\mu\, f_\rho \int\limits_0^1 du\,e^{-iu(p_\rho\,y)}
\phi_\parallel(u)
\ee
$\phi_{||}(z)$ is the meson distribution amplitude whose asymptotic form is
$\phi_{||}(z)=6z\bar z$, $f_\rho$ is the $\rho-$meson coupling constant,
$f_\rho=198\,\pm\,7\,$MeV. The generalized (skewed) quark distribution in
an unpolarized
nucleon target $F_\zeta$ is defined by the formula
\bea \label{F} &&\langle N(p_{2'})|\bar
q(0) \gamma^\mu q(y)| N(p_2) \rangle = \\ &&=\bar u(p_{2'})\gamma^\mu u(p_2)
\,\int\limits_0^1\,d\,x_1 \;\left[ e^{-ix_1(p_2y)} F^q_\zeta(x_1) -
e^{ix_2(p_2y)} F^{\bar
q}_\zeta(x_1)\right]\;. \nonumber
\eea
The necessary condition for the
validity of the collinear
factorization is the absence of pinch singularities in the integration over
the momentum
fraction $x_1$. The pinch contributions can produce terms which are
singular in the end-point
region of the momentum fraction $u$, e.g. terms behaving as $\sim
1/u,\;1/\bar u$ for $u, \bar
u \to 0$, so that an integral over $u$ would diverge.
For example, the result of an integral
with a pinch contribution leads to the result
\be
\label{pinch}
\int\frac{dx_1\;F_\zeta(x_1)}{[x_2+i\epsilon][x_2 + au-i\epsilon]} \;
\stackrel{u \to
0}{\longrightarrow}\; \sim \; -\frac{2i\pi F_\zeta(\zeta)}{au}\; \sim \;
\frac{1}{u}
\ee
whereas the one without pinch contribution gives
\be
\label{nopinch}
\int\frac{dx_1\;F_\zeta(x_1)}{[x_2+i\epsilon][x_2 + au +i\epsilon]}\;
\stackrel{u \to
0}{\longrightarrow} \; const.
\ee
The above example shows that in order to avoid pinch
singularities it is necessary to have the same $i\epsilon$ prescriptions in
the gluon poles
and the quark ones of diagrams in Fig.~(\ref{6d}) and that both poles in
expression
(\ref{nopinch}) contribute (see for details \cite{BISS}).


To show the absence of pinch singularities in the $x_1-$integration it is
then enough
 to examine  the
denominators appearing in all
diagrams in Fig.~(\ref{6d}). They
involve two gluonic
propagators which in a general kinematics, i.e. for arbitrary value of
the parameter $\alpha$, lead to denominators
\be
\label{k1}
k_1^2 + i\epsilon = -su\bar \alpha \left( x_1 - i\epsilon \right)
\ee
\be
\label{k2}
k_2^2 + i\epsilon = s\bar u\bar \alpha \left( x_2 + i\epsilon \right)\;.
\ee
and result in the factor of the integrand
\be
\label{prop}
\frac{1}{u \bar u ( x_1 - i\epsilon)( x_2 + i\epsilon)}\;.
\ee
As for the quark propagators which involve $x_{1,2}$ dependence,
 it is enough to
consider, for example, only those which appear in diagrams a and a':

\noindent diagram a:
\be
\label{a}
\left( zq_\rho - q - k_1 \right)^2 + i\epsilon = s(\bar \alpha \bar u + \bar z
\alpha)\left[x_2 +\bar u \,\frac{s_1 \bar z}{s(\bar \alpha \bar u + \bar z
\alpha)} + i \epsilon \right]
\ee
\noindent diagram a':
\be
\label{a'}
\left( zq_\rho - q - k_2 \right)^2 + i\epsilon = -s(\bar \alpha u + \bar z
\alpha
)\left[ x_1 -u \frac{s_1 \bar z}{s(\bar \alpha u + \bar z \alpha)} -i\epsilon
\right]
\ee
All remaining quark propagators with $x_{1,2}$ dependence either coincide
with
(\ref{a}) and (\ref{a'}) or they are obtained
 from
(\ref{a}) and (\ref{a'}) by the substitution $z \leftrightarrow \bar z$, which
is unimportant for the present discussion. A comparison of Eqs. (\ref{k1}),
(\ref{k2}),
(\ref{a}),
(\ref{a'}) leads to the conclusion that in each separate diagram in Fig.
(\ref{6d}) the $i\epsilon$ prescriptions in quark
and gluon propagators coincide and consequently there are no pinch
singularities.

\vskip.1in
{\large \bf 5.~~}Now we pass to the calculation of the process
(\ref{2mesongen}) in
the kinematics with large rapidity gap between $\rho^0_L$ and $\rho^+_L$, i.e.
characterized by conditions
\be
\label{cond}
s_1 \sim s \gg Q^2\,,\p^{\;2}\,,\;\;\;\;\;\zeta \sim 1\;.
\ee
They mean  that $\rho^0_L-$meson is produced in the fragmentation region of
the photon and consequently we should consider the limit
$\alpha \to 1$. Thus we shall neglect terms suppressed by
$\bar \alpha \to 0$ but  keeping the condition
(\ref{alphagap}).

In  such kinematics we can substitute the numerator of the gluon propagators in
the Feynman gauge by
\be
\label{g}
g^{\mu\,\nu} \to \frac{2}{s}\;p_1^\mu p_2^\nu \;,
\ee
where the Sudakov vector $p_2$ acts on the upper part of diagrams in
Fig.~\ref{6d}.

The gauge invariance of these diagrams allows us to omit
the component proportional to $q^\mu$ in the polarization vector
$\varepsilon_L^\mu$ of the longitudinaly polarized photon, i.e. to take it
 in the form
\be
\label{eL}
\varepsilon^\mu_L \to \frac{2Q}{s}\,p_2^\mu\;.
\ee
Calculations of diagrams in Fig.~\ref{6d} for quark $u$ contribution and
analogous
set of
diagrams for antiquark $\bar d$ contribution lead to the result
\bea
\label{sumM}
&&{\cal M}^{\gamma^*_L p \to \rho^0_L \rho^+_L n}= -
\frac{e\,\alpha_s^2\,\pi^2\,2^2\,C_F\,\sqrt{1-\zeta}\,
f_\rho^2}{4\,\sqrt{2}\,s^2\,N^2}\, \\
&&\int\limits_0^1 \;dz \,\phi_\parallel(z)\,
du \,\phi_\parallel(u)\,dx_1\,\frac{[F^u_\zeta(x_1) - F^{\bar
d}_\zeta(x_1)]}{[k_1^2 + i \epsilon][k_2^2 + i
\epsilon]}\,Tr\left(\hat p_\rho \, \hat p_1 \, \hat p_2 \, \hat p_1
\right)\,\sum\limits_{i=a,b,c,a',b',c'}\,I_i \;,\nonumber
\eea
with $C_F=\frac{N^2-1}{2\,N}$, and
\bea
\label{I}
&&I_a = Tr \left( \hat q_\rho\, \hat \varepsilon_L\, \frac{1}{z\hat q_\rho -
\hat q}\,
\hat p_2\, \frac{1}{z \hat q_\rho  -  \hat q -  \hat k_1}\, \hat p_2
\right) \;,
\nonumber \\
&&I_b = Tr \left( \hat q_\rho\, \hat p_2\, \frac{1}{z\hat q_\rho -
\hat k_2}\,\hat p_2\, \frac{1}{\hat q - \bar z \hat q_\rho}\, \hat
\varepsilon_L
\right) \;,
\nonumber \\
&&I_c = Tr \left( \hat q_\rho\, \hat p_2\, \frac{1}{z\hat q_\rho - \hat k_2}\,
\hat \varepsilon_L\, \frac{1}{ - \bar z \hat q_\rho + \hat k_1}\, \hat p_2
\right) \;,
\nonumber \\
&&I_{a'} = I_a(k_1 \leftrightarrow k_2)\,,\;\;\;\;\;
I_{b'} = I_b(k_1 \leftrightarrow k_2)\,,\;\;\;\;\;
I_{c'} = I_c(k_1 \leftrightarrow k_2)\;,
\eea
where we have as usual denoted $\hat v = \gamma^\mu\,v_\mu$ and we have
neglected all quark masses.

After calculation of the traces in Eq.~(\ref{sumM})
we can simplify the numerators of resulting  expressions  using conditions
(\ref{alphagap}) and keeping the denominators exact. In this
way we obtain that the scattering amplitude under study takes the form
\bea
\label{sumMmod}
&&{\cal M}^{\gamma^*_L p \to \rho^0_L \rho^+_L n}=
\frac{8\,e\,\alpha_s^2\,\pi^2\,C_F\,\sqrt{1-\zeta}\,Q\,s_1\,f_\rho^2}{\sqrt{2}\,
N^2\,s}
\nonumber \\
&&\int\limits_0^1 \;dz \,\phi_\parallel(z)\,
du \,\phi_\parallel(u)\,dx_1\,\frac{z \bar z [F^u_\zeta(x_1) - F^{\bar
d}_\zeta(x_1)]}{\bar \alpha^2 \,u\,\bar u\, [-x_1 + i \epsilon][x_2 + i
\epsilon]}\sum\limits_{i=a,b,c,a',b',c'}\,\tilde I_i \;,\nonumber
\eea
where
\bea
\label{Itilde}
&& \tilde I_a = \frac{\bar z}{z}\,\frac{1}{[Q^2\,(1-z\,\alpha)+\bar
\alpha\,z\,s_1]\,[x_2\,s\,(\bar u \,\bar \alpha + \bar z\,\alpha)+s_1\,\bar
u\,\bar z +i\epsilon]} \;,
\nonumber \\
&& \tilde I_{a'} = \frac{\bar z}{z}\,\frac{1}{[Q^2\,(1-z\,\alpha)+\bar
\alpha\,z\,s_1]\,[-x_1\,s(\bar \alpha\,u+\bar z\,\alpha)+s_1\,u\,\bar z
+i\epsilon]} \;, \nonumber \\
&& \tilde I_{c} = \frac{1}{[-x_1\,s(\bar z+\bar \alpha(u+z-1))+s_1\,u\,\bar z
+ i\epsilon]\,[s\,x_2(z+ \bar \alpha (1-u-z))+z\,\bar u\,s_1 + i\epsilon]}\;,
\nonumber \\
&&\tilde I_b = \tilde I_a(z \leftrightarrow \bar z)\,,\;\;\;\;
\tilde I_{b'} = \tilde I_{a'}(z \leftrightarrow \bar z)\,,\;\;\;\;
\tilde I_{c'} = \tilde I_c(z \leftrightarrow \bar z)\;.
\eea
Going to the limit (\ref{alphagap}) also in the denominators leads to
\bea
\label{aa'}
&&\tilde I_a + \tilde I_{a'}\;\to\;\frac{1}{Q^2\,z\,\bar z+z^2\,\p^{\;2}}\left[
\frac{1}{x_1\,s-s_1\,u+i\epsilon} +\frac{1}{-x_1\,s+s_1\,u+i\epsilon} \right]
\nonumber \\
&&= -2\,\pi\,i\,\delta(x_1\,s-s_1\,u) \;\frac{1}{Q^2\,z\,\bar
z+z^2\,\p^{\;2}}\;,
\eea
\be
\label{bb'}
\tilde I_b + \tilde I_{b'}\;\to \; -2\,\pi\,i\,\delta(x_1\,s-s_1\,u)
\;\frac{1}{Q^2\,z\,\bar z+\bar z^2\,\p^{\;2}}\;,
\ee
\be
\label{c}
\tilde I_c \;\to \; 2\,\pi\,i\,\delta(x_1\,s-s_1\,u)\;\frac{1}{Q^2\,z\,\bar
z+(1-u-z)^2\,\p^{\;2}}\;,
\ee
\be
\label{c'}
\tilde I_{c'} \;\to \; 2\,\pi\,i\,\delta(x_1\,s-s_1\,u)\;\frac{1}{Q^2\,z\,\bar
z+(u-z)^2\,\p^{\;2}}\;.
\ee
>From Eqs.~(\ref{aa'}), (\ref{bb'}), (\ref{c}) and (\ref{c'}) we see that
in our kinematics with a
large rapidity gap, the sum of all $\tilde I'$s contributions  leads to
a result which is purely
imaginary. Below we identify it with the discontinuity across the sum
of upper
parts of diagrams in Fig.~2.
Let us also observe that the delta function in the above expressions leads
to the
value \\
$x_1 = u\,s_1/s = u\,\zeta$, i.e. we probe the GPD's in the ERBL region.
Substituting (\ref{aa'}), (\ref{bb'}), (\ref{c}), (\ref{c'}) into
(\ref{sumMmod})
we obtain the result
\bea
\label{finalM}
&&{\cal M}^{\gamma^*_L p \to \rho^0_L \rho^+_L n}=
-\,\frac{i\,16\,s\,\zeta\,e\,\alpha_s^2\,\pi^3\,C_F\,\sqrt{1-\zeta}\,Q\,f_\rho^2
}
{\sqrt{2}\,N^2\,(\p^{\;2})^2}\;\int\limits_0^1\;
\frac{du\,\phi_\parallel(u)\,}{u^2 \,\bar u^2}\,\left[
 F^u_\zeta(u\,\zeta) - F^{\bar d}_\zeta(u\,\zeta)  \right]
\nonumber \\
&&\int\limits_0^1 \,dz\,\phi_\parallel(z)\,z\,\bar z\,\left(
\frac{1}{Q^2\,z\,\bar z+z^2\,\p^{\;2}}
+ \frac{1}{Q^2\,z\,\bar z+\bar z^2\,\p^{\;2}}
 - \frac{1}{Q^2\,z\,\bar z+(1-u-z)^2\,\p^{\;2}} \right.
\nonumber \\
&&\left. \hspace{9cm} - \frac{1}{Q^2\,z\,\bar z+(u-z)^2\,\p^{\;2}} \right)\;.
\eea
In the result (\ref{finalM})
all integrals  are well defined and converge, which shows that
the collinear factorization holds for our process. For its validity it is
important that
 the last integral over z in (\ref{finalM}) vanishes when $u$ or $\bar u
\to 0$.
This fact is not surprising since this integral is well known in BFKL
approach \cite{BFKL}
as the impact factor describing in the Born approximation the transition
$\gamma^*_L \to \rho^0_L$ by two gluon exchange (Fig.~(\ref{6d})).

The impact factor $J^{\gamma^*_L \to \rho^0_L}(\kt_1,\kt_2)$ depends on
the
transverse components of the momenta of the $t-$channel gluons $\kt_1$,
$\kt_2$,
$\kt_1+\kt_2=\p$.  It is defined as
the discontinuity
in the channel
($\gamma^*_L(q)\;g(k_1)$), i.e. in the Mandelstam
variable $s\beta \equiv (q+k_1)^2$
\be
\label{ifup}
i\,\delta^{a\,b}\;J^{\gamma^*_L \to \rho^0_L}(\kt_1,\kt_2)= \int
\frac{d(s\beta)}{2\pi i}
Disc_{s\beta}\frac{({\cal S}_{\mu\nu}^{\gamma^*_L \to \rho^0_L\,,a\,b}p_2^\mu
p_2^\nu)}{s^2}\;.
\ee
Here ${\cal S}_{\mu\nu}^{\gamma^*_L \to \rho^0_L\,,a\,b}$ is the S-matrix
element
for the above transition.
It contributes to the impact factor through its contraction with the light
cone components of the Sudakov vector $p_2$.
$a,\,b$ are the colour indices of the gluons.
In the Born approximation it is described by a
 quark loop with two $t$-channel gluons coupled
to it in all possible ways, i.e. by the upper parts of Fig. \ref{6d}.
The calculation of  $J^{\gamma^*_L \to
\rho^0_L}(\kt_1,\kt_2)$ is standard \cite{GI}, leading to the  result (for
nonvanishing quark mass $m_q$ case):
\be
\label{ifgamma}
J^{\gamma^*_L \to \rho^0_L}(\kt_1,\kt_2)=  -   f_\rho \frac{e
\alpha_s 2\pi
 Q}{N_c\sqrt{2}} \int\limits_0^1 dz\;z\bar z \phi_{||}(z)P(\kt_1,\kt_2)\;,
\ee
with
\bea
\label{P}
P(\kt_1,\kt_2=\p-\kt_1)=&&\frac{1}{z^2\p^{\;2}+m_q^2 +Q^2z\bar z} +
\frac{1}{{\bar z}^2\p^{\;2}+m_q^2 +Q^2z\bar z} \nonumber \\
&&-\frac{1}{(\kt_1-z\p\,)^2+m_q^2 +Q^2z\bar z} -
\frac{1}{(\kt_1-\bar z \p\,)^2+m_q^2 +Q^2z\bar z}\;.
\eea

Let us note also that the impact factor
$J^{\gamma^*_L \to \rho^0_L}(\kt_1,\kt_2)$ vanishes
linearly when
the transverse momentum of one of the
t-channel gluons, $\kt_1$ or $\kt_2$, vanishes. This fact is related to the
gauge
invariance of QCD. Its physical interpretation in this case is  that a
soft gluon cannot probe the structure of a colourless scattered object
(photon dissociating into a small quark dipole).

The last integral over z in Eq.~(\ref{finalM}) is proportional to
$J^{\gamma^*_L \to \rho^0_L}(\kt_1=u\p,\kt_2=\bar u \p)$ for $m_q=0$.
The  values appearing as arguments of this impact factor
follow in an obvious way from the conservation of transverse momentum in
the diagrams in Fig.~(\ref{6d}) and Eq.~(\ref{Sud}): the quark and
antiquark momenta
$x_1\,p_2$ and $x_2\,p_2$, respectively, are
longitudinal ones, consequently $\kt_1=-u\,\vec{p}_\rho=u\,\p$ and
$\kt_2= -\bar u \,\vec{p}_\rho= \bar u \,\p$.

Expressing the result (\ref{finalM}) in terms of the impact factor
$J^{\gamma^*_L \to \rho^0_L}$ we obtain
\bea
\label{CEN}
&&{\cal M}^{\gamma^*\,p\,\to \rho_L^0\, \rho^+_L\,n}=
i 8\pi^2 \zeta s \alpha_s f_\rho \sqrt{1-\zeta}
\frac{C_F}{N\,(\p^{\;2})^2}
\nonumber \\
&&\int\limits_0^1
\frac{\;du\;\phi_\parallel(u)}{ \,u^2 \bar u^2 }
 J^{\gamma^*_L \to \rho^0_L}(u\p,\bar u\p)
 \left[ F_\zeta^u(u\zeta)-F_\zeta^{\bar d}(u\zeta)\right]
\eea

Eq.~(\ref{CEN}) exhibits an interesting property which we would like to
emphasize. Our calculations were done consistently within the framework of the
collinear factorization. Nevertheless the result (\ref{CEN}) shows that
 the end-point singularities related to the
singular
behaviour of the hard t-channel two gluon propagators,
$\sim\; \frac{1}{u^2\bar u^2}$ for $u,\,\bar u\,\to 0$,
are regularized by both the $\rho$-meson distributions amplitudes
( $\phi_{||}(u)$) characteristic of the collinear
factorization
and by the impact factor
($J^{\gamma^*_L \to \rho^0_L}(u\p,\bar u\p)$) appearing naturally in the
BFKL approach. It would be very desirable to investigate
whether such
compensation of the end-point singularities
persists also after taking into
account  of radiative corrections.

\vskip.1in
{\large \bf 6.~~} All steps of the derivation which led us to Eq.~(\ref{CEN})
can be
now repeated for the process (\ref{2mesontr}) which involves the chiral-odd
transversity distribution whose investigation was the main motivation of our
studies.

To proceed with this program we need to define the corresponding
distribution amplitudes.
 We describe the production of a
transversely polarized $\rho$-meson  by means of its chiral-odd
light-cone distribution amplitude \cite{BalBr} defined
by the matrix element
\be
\label{rho+T}
\langle \rho_T(p_\rho,T) \mid \bar q(x) \sigma^{\mu \nu} q(-x)\mid 0
\rangle =i f_T \left(p_\rho^{\mu}\epsilon^{*\nu}_T -
p_\rho^{\nu}\epsilon^{*\mu}_T
\right)
\int\limits_0^1 du e^{-i(2u-1)(p_\rho x)}\;\phi_\perp(u)\;,
\ee
where the $\phi_\perp(u)=6u\bar u$ and $f_T(\mu)=160\pm10\,$MeV at the
scale
$\mu=1\,$GeV.

The generalized
(skewed) transversity distribution in nucleon target described by the
polarization vector $n^\mu$  is
defined by the formula
\bea
\label{FT}
&&\langle N(p_{2'},n)|\bar q(0) \sigma^{\mu\,\nu} q(y)| N(p_2,n) \rangle
= \\
&&=\bar u(p_{2'},n)\sigma^{\mu \,\nu} u(p_2,n)
\,\int\limits_0^1\,d\,x_1 \;\left[
e^{-ix_1(p_2y)}
F^{T\;q}_\zeta(x_1) - e^{ix_2(p_2y)} F^{T\;\bar q}_\zeta(x_1)\right]\;,
\nonumber
\eea
where $\sigma^{\mu\,\nu}=i/2[\gamma^\mu\,,\gamma^\nu]$.
Using the formulas  (\ref{rho+T}) and  (\ref{FT})
we obtain for the sum of the contributions of diagrams (\ref{6d}) the
expresion
\bea
\label{sumMT}
&&{\cal M}^{\gamma^*_L p \to \rho^0_L \rho^+_T n}= -
\frac{i\,e\,\alpha_s^2\,\pi^2\,2^2\,C_F\,\sqrt{1-\zeta}\,
f_\rho\, f_T}{64\,\sqrt{2}\,s^2\,N^2}\,
\\
&&\int\limits_0^1 \;dz \,\phi_\parallel(z)\,
du \,\phi_\perp(u)\,dx_1\,\frac{[F^{T\;u}_\zeta(x_1) - F^{T\;\bar
d}_\zeta(x_1)]}{[k_1^2 + i \epsilon][k_2^2 + i
\epsilon]} \,\sum\limits_{i=a,b,c,a',b',c'}\,I_i
\nonumber \\
&&Tr\left(\sigma^{\alpha \, \beta} \, \hat p_1 \, \sigma_{\mu\,\nu}
\, \hat p_1 \right) \,Tr\left(\sigma_{\alpha\,\beta}\,\hat
p_2\,\gamma^5\,\hat n
\right) \;\left(p_\rho^\mu\,\epsilon_T^{*\,\nu}-
p_\rho^\nu\,\epsilon_T^{*\,\mu}
\right)\nonumber
\eea
where $I_i$'s are given by Eq.~(\ref{I}) and $\gamma^5=\gamma_5=i\gamma^0
\gamma^1
\gamma^2 \gamma^3$. This formula is the analog of
Eq.~(\ref{sumM}). The product of traces in Eq.~(\ref{sumMT})
comes from the definitions of distributions in
Eqs.~(\ref{rho+T})
and
(\ref{FT}). This is the main difference between this expression
and
Eq.~(\ref{sumM}). The calculation of the above traces leads to the formula
\bea
\label{sumMTtrace}
&&{\cal M}^{\gamma^*_L p \to \rho^0_L \rho^+_T n}= - i\,\sin \theta
\frac{e\,\alpha_s^2\,\pi^2\,2\,C_F\,\sqrt{1-\zeta}\,
f_\rho \,\alpha\,s_1\,f_T}{\sqrt{2}\,s\,N^2}\,
\\
&&\int\limits_0^1 \;dz \,\phi_\parallel(z)\,
du \,\phi_\perp(u)\,dx_1\,\frac{[F^{T\;u}_\zeta(x_1) - F^{T\;\bar
d}_\zeta(x_1)]}{[k_1^2 + i \epsilon][k_2^2 + i
\epsilon]}\,\sum\limits_{i=a,b,c,a',b',c'}\,I_i \nonumber
\eea
where  $\theta$ is the angle between the transverse polarization vector of
the target
$\vec{n}$ and the polarization vector
$\vec{\epsilon}_T$ of the produced $\rho^+_T-$meson.
The comparison of Eq.~(\ref{sumMTtrace}) for transversity with the analogous
expression for $\rho^+_L-$meson production (\ref{sumM}) leads to the conclusion
that apart from obvious changes of coupling constants and distribution
amplitudes
the only difference between these two expressions is the presence in
(\ref{sumMTtrace}) of the additional factor $i\,\sin \theta$.
Thus the final result
for the scattering amplitude for the process
$\gamma^*_L p \to \rho^0_L \rho^+_T
n$ with transversity distribution takes the form
\bea
\label{CON}
&&{\cal M}^{\gamma^*\,p\,\to \rho_L^0\, \rho^+_T\,n}=
 -\,\sin \theta \;8\pi^2 \zeta s \alpha_s f_T \sqrt{1-\zeta}
\frac{C_F}{N\,(\p^{\;2})^2}
\nonumber \\
&&\int\limits_0^1
\frac{\;du\;\phi_\perp(u)}{ \,u^2 \bar u^2 }
 J^{\gamma^*_L \to \rho^0_L}(u\p,\bar u\p)
 \left[ F_\zeta^{T\;u}(u\zeta)-F_\zeta^{T\;\bar d}(u\zeta)\right]\;,
\eea
where $J^{\gamma^*_L \to \rho^0_L}$ is the same impact factor as in
(\ref{CEN}) given by Eqs.~(\ref{ifgamma}) and (\ref{P}).

\vskip.1in
{\large \bf 7.~~}
The simple form of the scattering amplitude  (\ref{CON}) (and of (\ref{CEN}))
involving the convolution of the impact factor, the vector meson
distribution amplitude and the
GPD suggests a search for other processes in which the transversity can be
described in
a similar framework and which are may be  easier for experimental
studies. If
for such a new process the impact factor of corresponding
transition is known it is enough to replace in Eq.~(\ref{CON}) the
impact factor $J^{\gamma^*_L \to
\rho^0_L}$ by another one to obtain the appropriate scattering amplitude.

The most obvious candidate is the process (\ref{2mesontr}) with the
transversely polarized
photon \\
$\gamma^*_T \;\; N  \to  \rho_L^0\;\; \rho_T^+ \;\;N'$. The impact
factor $J^{\gamma^*_T \to \rho^0_L}$ has then the
form:
\be
\label{TL}
J^{\gamma^*_T \to \rho^0_L}(\kt_1,\kt_2=\p-\kt_1)=
-\frac{e\,\alpha_s\,\pi\,f_\rho}{\sqrt{2}\,N}\;\int\limits_0^1\,dz\,(2z-1)\,
\phi_\parallel(z)\,\left( \vec{\varepsilon}\,\vec Q_P \right)\;.
\ee
with $\vec{\varepsilon}$ being the polarization vector of the initial photon,
and
\bea
\label{Q}
&&\vec Q_P(\kt_1,\kt_2=\p-\kt_1)=
\frac{z\,\p}{z^2\,\p^2+Q^2\,z\,\bar z +m_q^2}
- \frac{\bar z\,\p}{\bar z^2\,\p^2+Q^2\,z\,\bar z +m_q^2}
 \\
&&\hspace{3cm}+\frac{\kt_1 - z\,\p}{(\kt_1 - z\,\p)^2+Q^2\,z\,\bar z +m_q^2}
-\frac{\kt_1 - \bar z\,\p}{(\kt_1 - \bar z\,\p)^2+Q^2\,z\,\bar z +m_q^2}\;.
\nonumber
\eea

Another example is the process (\ref{2mesontr}) with $\rho^0$ replaced by
heavy $J/\Psi-$meson. One could study either $\gamma^*_L \;\; N  \to
J/\Psi_L\;\; \rho_T^+\;\;   N'$ involving the impact factor
\bea
\label{charmLL}
&&J^{\gamma^*_L \to
J/\Psi_L}(\kt_1,\kt_2=\p-\kt_1)=
\\
&&=-\frac{8\pi\,e\,\alpha_s\,Q\,f_{J/\Psi}}{3\,N}\;
\left(\frac{1}{\p^{\;2}+Q^2+4\,m_c^2} - \frac{1}{(2\,\kt_1 -\p)^2 +Q^2+
4\,m_c^2}
\right) \;, \nonumber
\eea
or the process $\gamma^*_T \;\; N  \to
J/\Psi_T\;\; \rho_T^+\;\;   N'$ for which the impact factor $J^{\gamma^*_T \to
J/\Psi_T}$ has the form
\bea
\label{charmTT}
&&J^{\gamma^*_T \to
J/\Psi_T}(\kt_1,\kt_2=\p-\kt_1)=
\\
&&=\frac{4\,e\,\alpha_s\,\pi\,m_c\,f_{J/\Psi}}{3\,N}\;(\vec \varepsilon
\,\vec \epsilon^{\;*})\,\left( \frac{1}{\p^{\;2} + Q^2 +4\,m_c^2} -
\frac{1}{(2\,\kt_1-\p)^2 + Q^2 +4\,m_c^2}   \right)\;.
\nonumber
\eea
These impact factors are obtained from the corresponding ones for light
quarks by applying a standard non-relativistic approximation for
the $J/\Psi$-meson vertex, i.e.
by approximating
the distribution amplitude  by
$\phi_{j/\Psi}(z)=\delta(z-1/2)$.
The coupling constant $f_{J/\Psi}^2 =\frac{27\,m_{J/\Psi}\,\Gamma_{J/\Psi
\to e^+
e^-}}{16\,\pi\,\alpha_{em}^2}$ is expressed in terms of the
width $\Gamma_{J/\Psi \to e^+\,e^-}$
and
$\alpha_{em}={e^2}/{4\pi}$.
This last process (\ref{charmTT}) can be of course also studied both for
virtual as well as
real photon.

One can also study the transversity in the inelastic DVCS: $\gamma^*_{L/T}
\;\; N  \to \gamma\;\; \rho_T^+\;\;   N'$. The expressions for the
corresponding impact factors
$J^{\gamma^*_T \to \gamma_{T'}}$ and
$J^{\gamma^*_L \to \gamma_{T'}}$
  are also known and can be found e.g. in
\cite{BalKu} (
Eq.~(11)  and Eq.~(12), respectively).

Another very natural and interesting choice is the
hadroproduction of a vector meson in such "asymmetric" kinematics.
Hadron collider experiments should in principle give
a powerful  opportunity to test our factorization scheme and give access
to transversity. Diffractive physics is indeed a topical subject of
investigation of the Tevatron at FNAL, RHIC at BNL and LHC at
CERN facilities and high quality data will be obtained in the near future
at these places.  Contrary to the $\gamma^*\,{\cal P}\,\rho-$coupling, the
$p\,{\cal P}\,p-$coupling is not yet under theoretical control
(diffractive factorization breaking in
hadron hadron collisions\cite {nonfact} may forbid the extension of the
electroproduction case; moreover, one should take into
 account the Landshoff mechanism \cite{Land}), and deserves much more
study before we can apply the ideas of the present paper to meson production in
diffractive $p\,p$ scattering with asymmetric kinematics.

\vskip.1in
{\large \bf 8.~~}
In conclusion, the electroproduction process of two mesons with large
rapidity gap can be described consistently within the collinear factorization
approach.  Higher order studies are
necessary  to establish its validity beyond the Born order. On the other
hand the chiral-odd GPD may now be accessed in a feasible reaction, namely
$\gamma^*\,p\;\to \;\rho^0_L\,\rho^+_T\,n$.
An estimate of the cross-section of this reaction requires a knowledge of the
chiral-odd GPD. No model has yet been proposed for this quantity.
We believe that further
improvement
of the theoretical understanding of hadronic impact factors may help us to
access
this chiral-odd GPD also in hadronic diffractive reactions at RHIC and LHC.

\vspace*{1cm}
{\Large \bf Acknowledgments}
\vskip.1in
We acknowledge useful discussions with M.~Diehl, H.G.~Dosch, S.~Munier,
D.~M{\"u}ller, A.~Sch{\"a}fer and
G.~Sterman.
D.I.,  L.Sz. and O.T. acknowledge the warm hospitality in {\'E}cole
Polytechnique.
 This work is supported in part
by the TMR and IHRP Programmes of the European Union, Contracts
No.~FMRX-CT98-0194 and No.~HPRN-CT-2000-00130,
 INTAS Project 587 (Call 2000), Deutsche Forschungsgemeinschaft DFG
(Hadronische Spinphysik, Scha-458) and BMBF (06OR984).

\end{document}